\documentclass[letterpaper,11pt]{article}
\usepackage[dvips]{graphicx}
\title{Realization of a Laser Cooled Microwave Power Standard} 
\author{David C. Paulusse, Nelson L. Rowell, and Alain Michaud\footnote{The authors are with the Institute for National Measurement Standards, National Research Council, Ottawa, ON., Canada, K1A 0R6. email: Alain.Michaud@nrc-cnrc.gc.ca. This paper was presented at the Conference on Precision Electromagnetic Measurements (CPEM 2002), Ottawa, ON., Canada, 16-21 June, 2002. }}

\date{June 18, 2002}

\begin{document}

\maketitle

\paragraph{Abstract--}
We demonstrate the feasibility of a novel microwave power standard based on the electromagnetic interaction with laser-cooled atoms. Under the effect of the radiation, the internal state populations undergo a Rabi flopping oscillation. Measurement of the oscillation frequency allows the determination of the electromagnetic field strength. The measurements are made over a dynamic range of 18 dB and the standard deviation of the measurements is about 0.3\%. The field strength was linear with the Rabi frequency within the accuracy of our measurements.\\
 


\begin{centering}
\section*{\sc Introduction}
\end{centering}

SINCE the application in the early eighties of concepts related to mechanical effects of light the field now known as laser cooling and trapping has become one of the more studied in experimental physics. Reference \cite{rice} gives an index of known links related to quantum optics and laser cooling. It is updated regularly. The first successful applications were in the fields of metrology and spectroscopy, where the atoms in an atomic vapour were probed using laser beams. With these methods the interaction time was increased by many orders of magnitude. Barely a decade after the initial experiments, the absolute measurement of the basic unit of time, the second, has been improved by a factor of five with cooling methods. The new techniques are also expected to improve precision measurements of physical effects such as gravity, magnetic field, rotation of earth, and fundamental constants.

This paper summarizes our progress towards using the electromagnetic interaction as a tool in the field of microwave power measurements. To this end we have built a new type of apparatus in which laser cooled Rubidium atoms are subjected to the electromagnetic field. Under the effect of the radiation, the internal state populations undergo a Rabi oscillation. The measurement of the Rabi frequency allows the determination of the electromagnetic field's strength.

In the following, we give a short description of the apparatus, we show how it was used to measure the relative power of a microwave signal, and we discuss its resolution and how it could be upgraded to provide an absolute standard for microwave power.\\

\begin{centering}
\section*{\sc II. Atomic Power Standard}
\end{centering}

Although the standard practical way to measure microwave power is the calorimetric technique, several other approaches have been employed \cite{fantom}. In one such technique an atomic resonance was used to monitor saturation in the absorption spectrum of a gas (NH$_{3}$) filling a long section (7~m) of waveguide (\cite{hashimoto}, \cite{fantom} section 10.10). When the frequency of the radiation was swept through the resonance (24 GHz), the power could be deduced from the line shape with knowledge of parameters such as temperature and gas pressure (relaxation constant). \\

\subsubsection*{A. Principle of Operation}

Our experiment differs from the one described above because with laser cooling the atoms are not perturbed significantly by collisions with the walls or other atoms. Furthermore since the velocity of the atoms is only a few mm/s, the interaction time is greatly increased to the point that it is now limited only by the effect of gravity. Figure 1 shows a schematic of the experimental setup. A typical measurement involves cycling repeatedly through the three steps described below: \\

\begin{figure}
\includegraphics[height=6in]{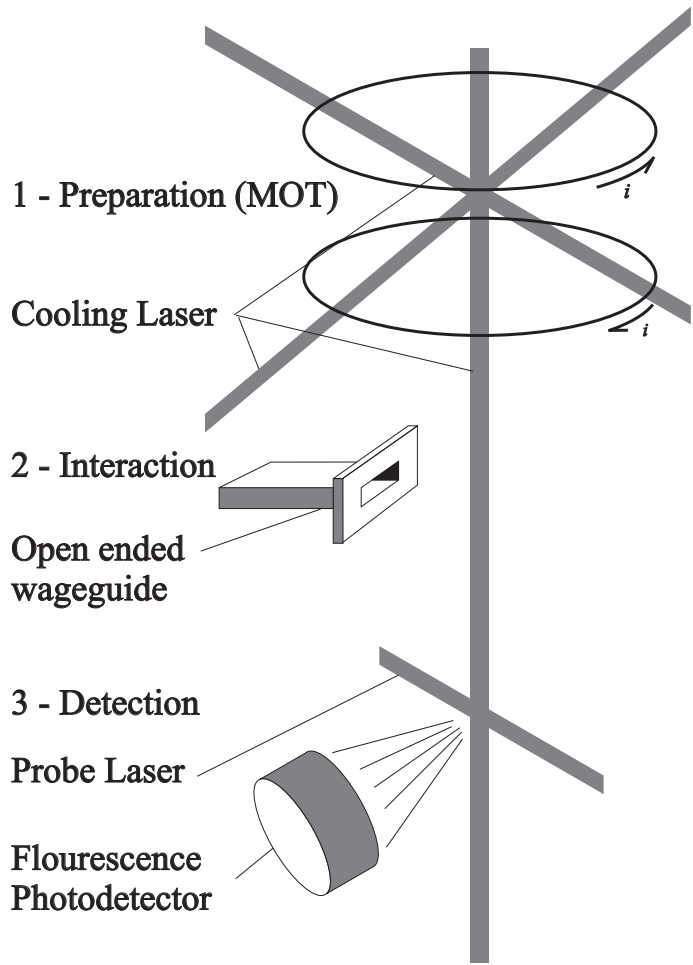}
\caption{Schematic of the experimental system (not to scale). In "1", the atoms are trapped in the intersection of the laser beams, then they fall in front of the waveguide "2" and in the detection zone "3". The glass vacuum chamber surrounding the vertical laser beam is not shown.}
\end{figure}

\subsubsection*{1) Preparation of a cold atom sample}

First, approximately 10$^{8}$ atoms of Rb$^{87}$ are captured in a standard magneto-optical trap (MOT) \cite{metcalf}. In essence, three orthogonal counter-propagating laser beams cross at the center of a pair of Anti-Helmoltz coils. The operating conditions are, (i) The "cooling laser" must be detuned slightly (-10 MHz) from the 5s$_{1/2}$, F=2 to 5p$_{3/2}$, F'=3, atomic transition, (ii) The "repumping laser" must be tuned exactly to the wavelength of the 5s$_{1/2}$, F=1 to 5p$_{3/2}$, F'=2 transition, (iii) Proper laser beam polarization conditions must be set, and (iv) An adequate gradient of magnetic flux density (about 1 mT/cm) must be present. 

Next the magnetic field is ramped to zero, the cooling laser is detuned, and the light intensity reduced. This results in a lower temperature sample which is called "optical molasses" (\cite{metcalf} chap. 7). 
Subsequently a short pulse of resonant light can be applied to optically pump the atoms in the 5s$_{1/2}$, F=2 state. Instead of doing this, however, we use a simpler but less efficient expedient, that of simply turning off the repumping light earlier then the cooling laser. 
\\

\subsubsection*{2) Interaction with the microwave field}

After the laser beams has been turned off and the magnetic field nulled, the atoms fall in the interaction zone. At this point all the atoms are in the F=1 state. 

In this experiment, we apply a resonant pulse of a known width and we measure the population of the level 5S$_{1/2}$, F=2 at the end of the pulse. We repeat the experiment varying the pulse length each time. It is therefore possible to deduce the value of from by evaluating the period of the relative population of the F=2 state. 

In the following analysis, we make the assumptions that there are only two levels, 5S$_{1/2}$, F=1 and 5S$_{1/2}$, F=2, their lifetimes are longer than the interaction time, and the electromagnetic field is resonant with the transition. Under the influence of the field, the populations of both levels oscillate as described in the following equations: \\

\begin{equation}
 P_{1}(t) = \cos^{2} (bt)
\end{equation}

\begin{equation}
 P_{2}(t) = \sin^{2} (bt)
\end{equation}

Where $P_{1}(t)$ and $P_{2}(t)$ are the relative populations of the states, F=1 and F=2 and the Rabi frequency, $b$ is proportional to the magnetic field amplitude. We also assume that the frequency of the transition is much larger than $b$, otherwise a more complex model is necessary to describe the mechanism \cite{silverman}. \\

\begin{figure}[p]
\includegraphics[width=5in]{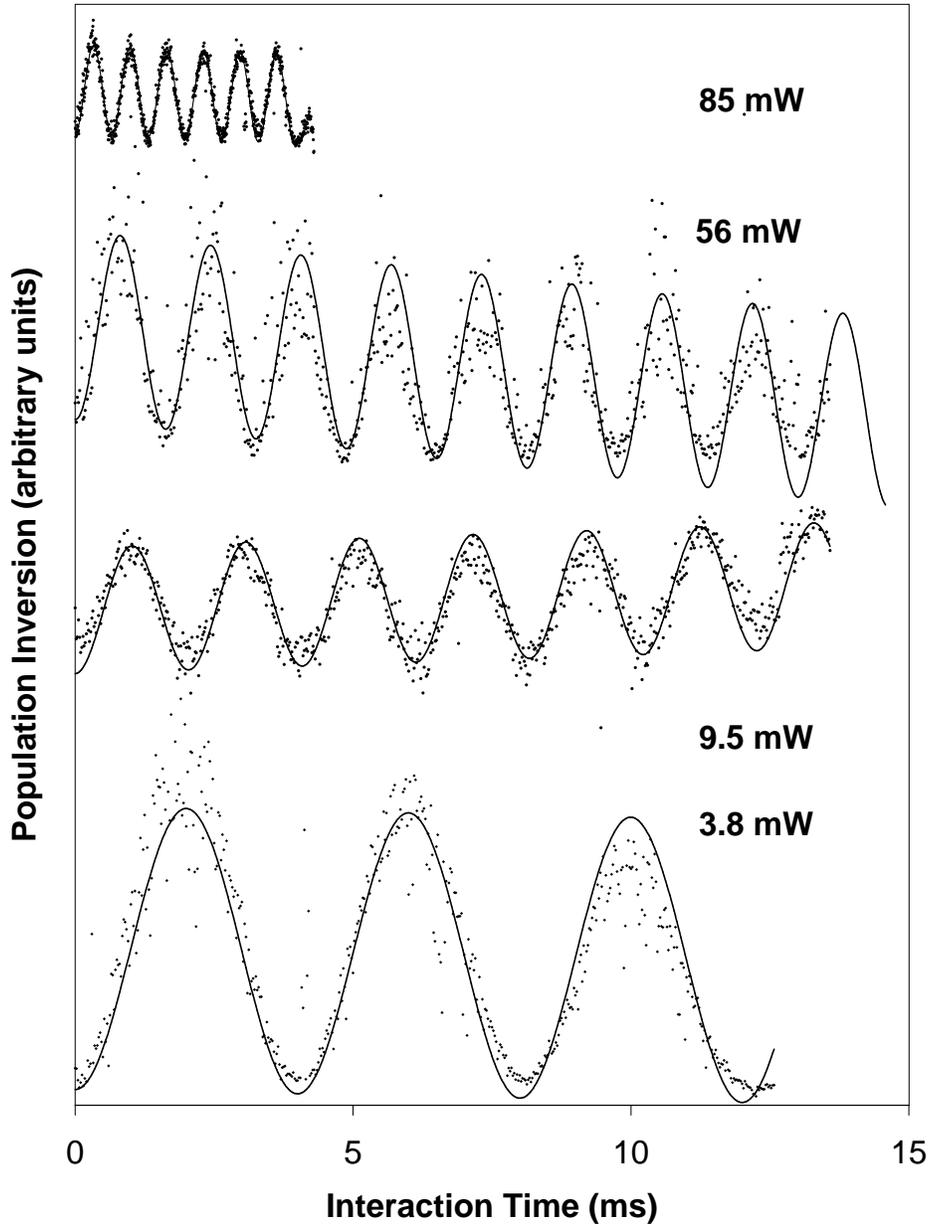}
\caption{Time evolution of the detected sample population inversion. The parameter displayed at the right, is the power that was sent to the waveguide. For the upper three curves, an offset has been added to the ordinate in order to allow plotting the curves on the same graph. For the lower curve (3.8 mW), the horizontal axis represents the actual zero and one can observe the good contrast on these fringes.}
\end{figure}

\subsubsection*{3) Detection of population inversion}

After the atoms have fallen through the interaction zone they are probed using a laser beam. There are many techniques to probe the atoms but the one that we use is the simplest one; i.e. a single resonant laser beam (5S$_{1/2}$, F=2 to 5P$_{3/2}$, F'=3 transition) in the trajectory of the atoms. As the atoms fall through this laser beam, their fluorescence light is collected, focused on a photodetector, and the measured intensity is recorded. The amplitude of this time-of-flight signal is proportional to the population of the level 5S$_{1/2}$, F=2 and the total number of atoms in the cloud. A more Efficient system would use a second laser to probe the atoms in the 5S$_{1/2}$, F=1 level as well. As this would be a more complicated detection technique to perform, it has not yet been implemented despite the anticipated improvements in signal to noise ratio. \\

\subsubsection*{A. Description of the Apparatus}

\subsubsection*{1) Laser system}

The laser system is based on three laser diodes. The cooling laser is kept in resonance with 5S$_{1/2}$, F= 2 to 5P$_{3/2}$, F=3 transition. The output of this laser injects a second laser used as an amplifier. The output power that is sent to the vacuum chamber is approximately 35 mW. 

The repumping laser, similar to the first one but tuned to the F=1 to the F'=0 is necessary to prevent optical pumping. Its characteristics (power and linewidth) are not critical. 

\subsubsection*{2) Microwave Field}

The geometry of the radiating system is very critical. It has to be simple and symmetrical so that the field distribution can be calculated. Moreover it must have sufficient optical access for the many laser beams. To inject the microwave field we have used an open-ended waveguide (R-70, 6.8345~GHz). This device has been studied \cite{gardiol} and its simple geometry will allow the construction of a mechanically precise device. 

The microwave signal is generated by a synthesizer (Agilent 83620A) followed by a pulse modulator (Agilent 11720A) and an amplifier (Agilent 8349B). The output power from the amplifier is measured using a thermistor mount, through a 9.5 dB directional coupler and a NBS type IV bridge (Tech USA) and a DC voltmeter (Agilent 3457A). The pulse modulator is turned on for a few seconds and, to compensate for systematic effects such as drift, the power is measured before and after an experiment is run. 

We should mention here that the measurement of a CW signal is in principle possible with our system. This would, however, require two laser pulses that are spaced by a time interval. The first pulse would perform the optical pumping as described previously. The mechanical effect of the second pulse (F=1 to F'=0) would blow away the atoms that are left in the F=1 state after a certain time. Following this sequence, the total number of atoms as a function of the time interval would show the Rabi fringes. Although this system has yet to be tested, it would allow the measurement of continuous wave signals.\\
\subsubsection*{3) Vacuum chamber}

The vacuum chamber consists in a glass cube (50 mm x 50 mm x 90 mm) attached to a high vacuum pump and a reservoir filled with rubidium metal. The laser beams, and microwave field are directed onto the atoms through the optical-quality glass plates forming the walls of the cube.
To evaluate accuracy as a standard, the dielectric loading of the glass will have to be evaluated or alternatively, a new system could be built, one in which the waveguide penetrates into the vacuum chamber. For these reasons we have not attempted to evaluate the absolute accuracy of this standard. Instead the experimental results in this paper focus mainly on the resolution and linearity of the system when used as a microwave detector.\\
%
\section*{\sc III. Experimental Results and Discussions }

\subsubsection*{A. Rabi Frequency}

Figure 2 shows four time-evolution plots at different power levels. The height of each dot is proportional to the number of atoms in the F=2 state. The solid curve is derived from a cosine function plus a linear baseline. By virtue of the laser cooling, the interaction time can be as long as 20 ms limited only by the effect of gravity. Also, due to absence of collisions in the cooled atoms, the fringes do not fade significantly, as would be the case for atoms at room temperature.

The noise on those curves is due to the shot-to-shot fluctuation of the total number of atoms in the sample. Such fluctuation, common with MOT's, cannot be reduced significantly. However, the system could be improved by implementing a detection technique in which the readings would be normalized by the total number of atoms in the sample using the two detection laser method (described above). 

Nevertheless even if the time evolution plots are moderately noisy, the frequency can be estimated with relative accuracy due to the periodic structure of the curves. This shows the potential for this technique. Also we see that any offsets or tilts caused by laser frequency drift or variation in detector performance have little effect on the estimated frequency. 

Each measurement cycle lasts about 3 seconds resulting in a total measurement time of about 30 minutes for each plot in Fig. 2. The measurement time could be reduced by increasing the background gas density, which depends exponentially on the oven temperature. However this was not found be necessary and the oven was set to room temperature. 

Also, the complete time-scans are plotted for illustrative purpose only. If the power were known approximately, then the measurement of certain specific points for every cycle would suffice for an accurate frequency estimation. Finally, long-term drift could be measured easily by continuously monitoring the position of the last fringe. \\

\subsubsection*{B. Linearity}

We made several time-scans like the ones shown in Fig. 2, to cover a dynamic range of about 20 dB. The high power end was limited by the maximum power of the source whereas the lowest measurable power was limited by the increasingly long interaction times at low power. In power the range was between 1.5 and 85 mW. A summary of these measurements is plotted in Fig. 3(a). As expected the Rabi frequency is proportional to the field amplitude. 

\begin{figure}[p]

\vspace{-1in}

\includegraphics[width=5in]{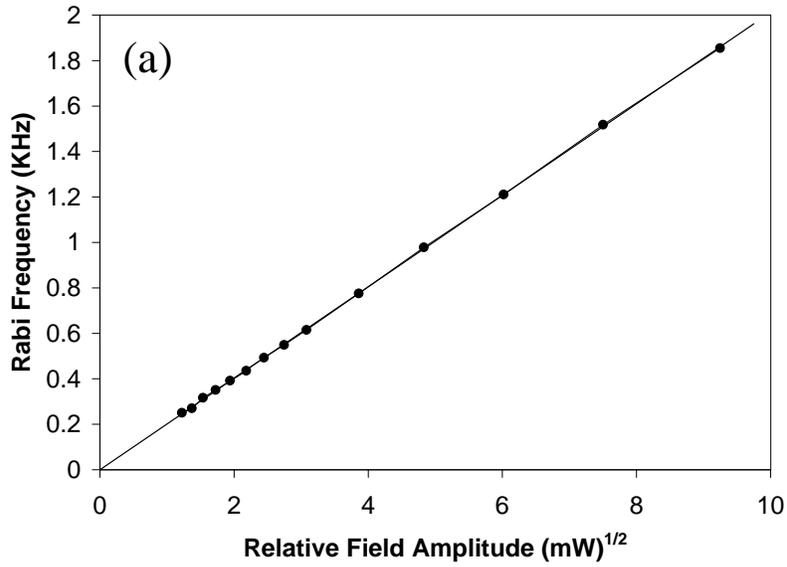}

\includegraphics[width=5in]{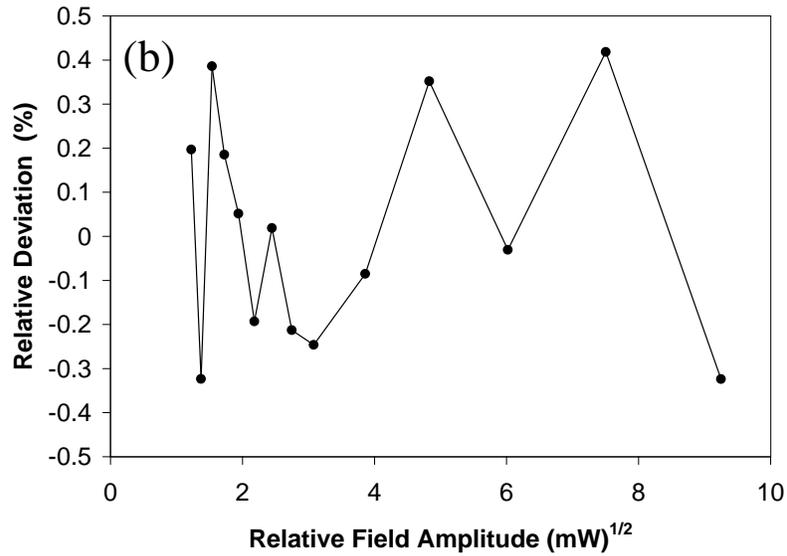}

\caption{ (a: upper) Measured Rabi frequency as a function of the relative field amplitude. The gain of the antenna is not known, but for ease of comparison with figure 2, the horizontal axis units are expressed as a function of the square root of the power sent to the waveguide. (b: lower) Deviation from a straight line in (a) normalised by the value of the highest point.}

\end{figure}

Figure 3 (b) shows the relative deviation of the points in Fig. 3(a) from a straight line. It shows that the typical noise on the measurements is about ± 0.3\%. Also we see that, if there is any non-linear dependence, it is smaller than the random noise in the measurements. More precision in the measurement over a wider power range will be necessary to verify the linearity of this device to a higher degree of certainty. \\

\begin{centering}
\section*{\sc IV. Conclusion and Outlook}
\end{centering}

We have demonstrated that the precise measurement of the field strength can be done using the electromagnetic interaction with cold atoms. The standard deviation of the measurements was about 0.3\%. The measurements were made over a dynamic range of 18 dB. Under those conditions, Rabi frequency was observed to be linear with microwave field strength. 

Further investigations and improvements stimulated by the present work include: 

Improvements to the system's design to increase the signal-to-noise ratio, dynamic range, to allow CW operation, and to eliminate any field perturbations in the interaction zone.

A systematic theoretical evaluation considering both the radiation pattern and the magnetic susceptibility of the atomic sample. \\

\begin{centering}
\section*{\sc Acknowledgement}
\end{centering}

We thank the following people for their contributions: J.~Alcock, A.~Avery, S.~Cundy, P.~Filipski, C.~Jacques, D.~Scobie, D.~Singleton and B.~Wood. \\


\begin{thebibliography}{9}

\bibitem{rice} \textsc{D.~Rice}. "Quantum Optics and Atom Optics links" [Online], Available: http://www.quantumoptics.net

\bibitem{fantom} \textsc{A.~Fantom}. \emph{Radio Frequency and Microwave Power Measurement},	 \'Ed. Peter Peregrinius Ltd., London, 278 pp.,  1990, IEE press, Available: http://www.iee.org

\bibitem{hashimoto} \textsc{S.~Hashimoto, M.~Yamaghushi}, "Microwave Power Measurement Using Absorption Spectrum of Ammonia Gas", \emph{IEEE Trans. Instrum. Meas.}, {\bf IM-25}, pp. 348-352, Dec. 1976.

\bibitem{metcalf} \textsc{H.J.~Metcalf, P.~van~der~Straten}, \emph{Laser Cooling and Trapping}	New York, Springer-Verlag, 1999, pp. 156–162.

\bibitem{silverman} \textsc{M.P.~Silverman}, \emph{Probing the Atom}, Princeton University Press, Princeton, N.J., 2000, 226 pp.

\bibitem{gardiol} \textsc{F.E.~Gardiol}, "Open-ended waveguides, principles and operations", in \emph{Advances in Electronics and Electron Physics}, P.E. Hawkes, Ed., Academic Press, New York, 1985, pp. 139-187.

\end{thebibliography}
\end{document}